# A low-temperature scanning probe microscopy system with molecular beam epitaxy and optical access

Ze-Bin Wu, Zhao-Yan Gao, Xi-Ya Chen, Yu-Qing Xing, Huan Yang, Geng Li, Ruisong Ma, Aiwei Wang, Jiahao Yan, Chengmin Shen, Shixuan Du, Qing Huan,[†] and Hong-Jun Gao [†]

*Institute of Physics & University of Chinese Academy of Sciences, Chinese Academy of Sciences, Beijing 100190, China*

*Key Laboratory for Vacuum Physics, Chinese Academy of Sciences, Beijing 100049, China*

A low-temperature (LT) ultra-high vacuum (UHV) scanning probe microscopy (SPM) system with molecular beam epitaxy (MBE) capability and optical access was conceived, built, and tested in our lab. The design of the whole system is discussed here, with special emphasis on some critical parts. We made an SPM scanner head with a modified Pan-type design, enclosed by a double-layer cold room under a bath type cryostat. The scanner head is very rigid, compatible with optical access paths, and can accommodate both scanning tunneling microscope (STM) tips and atomic force sensors. Two piezo-actuated focus-lens stages are mounted on the two sides of the cold room to couple light in and out. To demonstrate the system's performance, we performed STM and scanning tunneling spectroscopy (STS) studies. The herringbone reconstruction and atomic structure of Au(111) surface were clearly resolved. The *dI/dV* spectra of an Au(111) surface were obtained at 5 K. In addition, a periodic 2D tellurium (Te) structure was grown on Au(111) surface using MBE.

---

[†]Author to whom correspondence should be addressed. Electronic mail: huanq@iphy.ac.cn, hjgao@iphy.ac.cn

INTRODUCTION

Scanning probe microscopy (SPM)[1-4] is a powerful tool in surface science due to its outstanding spatial resolution. The combination of SPM with molecular beam epitaxy (MBE) in a single system enables *in situ* preparation and characterization of materials.[5] Many interesting 2D materials such as germanene,[6] $PtSe_2$,[7] CuSe,[8] $MoSe_2$[9] and $PdTe_2$,[10] and heterostructures such as $HfTe_3/HfTe_5/Hf$[11] and $MoSe_2/HfSe_2$[12] have been prepared by MBE and then investigated *in situ* by SPM. Besides, the introduction of light into SPM can further extends its applications. Based on this, many cutting-edge techniques emerge, such as tip-enhanced Raman spectroscopy (TERS),[13-17] ultrafast-laser-coupled scanning tunneling microscopy (STM)[18, 19] and tip-induced photoluminescence,[20] etc.[21, 22] In particular, the combination of SPM and ultrafast-laser systems enables scientists to study materials at high resolution, both spatial and temporal.

In this paper, we report the construction and the performance testing of a newly developed low-temperature (LT) SPM system with MBE and two optical access paths. The custom-made SPM scanner head features a modular design with a unibody titanium frame. The tip holder accommodates both scanning tunneling microscope (STM) tips and atomic force sensors, providing the system additional flexibility. A double-layer cold room is mounted at the bottom of the cryostat, and two piezo-driven focus lenses are mounted on the two sides. The vacuum level in the SPM chamber and the MBE chamber extends below $5 \times 10^{-11}$ torr after baking. The sample temperature is around 5 K after cooling down. The performance of the system is demonstrated by the successful growth and characterization of a 2D tellurium structure using MBE and STM. The system exhibits good stability and compatibility, and thus proves to be a versatile and powerful tool for growth and characterization of low-dimensional materials.

I. SYSTEM DESIGN

A. Whole structure

Figures 1 (a) and (b) show a 3D model and a photograph of the system. The whole system is fixed on a frame of aluminum beams (beam cross section size is 80

mm×80 mm) supported by four pneumatic vibration isolators. The estimated mass center of the system is indicated by a red dot in Fig. 1 (a). The system has four main chambers – an SPM chamber, an MBE chamber, a preparation chamber, and a load-lock chamber. A 700 L/s turbomolecular pump is installed on the MBE chamber and serves as the main pump for the system. Another 300 L/s turbomolecular pump is shared by the preparation chamber and the load-lock chamber. The outlet ports of both pumps are connected to a buffer chamber equipped with an 80 L/s turbomolecular pump. This design ensures a better vacuum level of the main chambers. The SPM chamber, the MBE chamber, and the preparation chamber are equipped with ion pumps of 300 L/s, 300 L/s, and 75 L/s, respectively. The internal volume of the load-lock chamber is only 195 mL, ensuring a short pump-down time from atmospheric pressure to high vacuum. Before baking, all the chambers and bakable parts of the system are enclosed in a tent made of thermal isolation materials. Two 5 kW heaters with fans are used to heat up the system to around 150 °C. Both the SPM chamber and the MBE chamber could reach below $5 \times 10^{-11}$ torr after one week's baking.

Samples and tips are mounted on a standard flag-style sample holder. Typically, the sample is loaded into the system from the load-lock chamber and then transferred to the sample stage of the manipulator vertically mounted on the preparation chamber by the transfer rod in the load-lock chamber. Cycles of ion bombardment and annealing (up to 1400 K) of the sample can be performed in the preparation chamber. After the cleaning process, the sample is transferred to the other manipulator, horizontally mounted on the MBE chamber, by the transfer rod in the preparation chamber. This two-transfer-rod design prevents direct connection between the load-lock chamber and the MBE chamber, helping to maintain a better vacuum inside the MBE chamber where the desired material is grown. After growth, the sample is transferred to the SPM chamber by the horizontal manipulator. Inside the SPM chamber, a wobble stick is used to transfer the sample from the manipulator to the scanner head for cooling down and imaging. The SPM tip transfer follows the same path.

The MBE chamber, as shown in Fig. 2, can accommodate six evaporators simultaneously. As mentioned above, a horizontal manipulator is mounted in the chamber, on which the sample can be heated up to 1400 K by electron beam heating or cooled down to 120 K by continuous flow of liquid nitrogen. A cryo-shroud made of oxygen-free high-conductivity copper is installed close to the inner chamber wall and attached to a small cryogen tank of ~ 2.3 L hung underneath the top flange of the chamber through two stainless steel pipes. This cryo-shroud works as a cryopump and helps to improve the vacuum level. Moreover, 15 KeV Reflection High-Energy Electron Diffraction (RHEED) equipment is installed for real-time monitoring of layered material growth. After deposition of materials by MBE, the sample can be annealed *in situ* on the manipulator.

**B. Cold room**

A customized commercial cryostat from CryoVac GmbH is mounted on top of the SPM chamber, as indicated in Fig. 1 (b). The volumes of the inner and outer cryogenic vessels are 4 L and 17 L for liquid helium (LHe) and liquid nitrogen ($LN_2$), respectively. A double-layer cold room, consisting of a $LN_2$ shield and a LHe shield, is mounted at the bottom of the cryostat. The $LN_2$ shield is composed of polished aluminum sheets attached to an octagonal frame of eight aluminum columns, and the LHe shield is similar but has a cuboid frame of four aluminum columns, as shown in Fig. 3 (a). The aluminum frames are good thermal conductors, ensuring fast and efficient cooling of the shields. Inside the LHe shield, the SPM scanner head is suspended by three mechanical springs made of Inconel wire for vibration isolation. The calculated natural frequency of this suspension-spring system is around 1.56 Hz. Eight SmCo magnets are installed around the scanner head for vibration damping in both lateral and vertical directions. The combination of the mechanical springs and magnetic damping proves to be efficient and is widely used in different kinds of SPM systems.[23-25]

Photoassisted SPM can provide a wealth of information of physical properties with atomic or molecular resolution. Quite a few approaches have been developed to build SPMs with optical access paths.[26-29] In our system, the optical paths inside the SPM chamber are shown in Fig. 3 (b). The angle of incidence relative to the normal of the sample surface is 50°. A piece of plane-convex fused silica lenses of 25.4 mm in diameter is placed on each of the two sides of the scanner head, located 60 mm away from the tip-sample junction. We use a pair of piezo stack actuators, fastened to the $LN_2$ shield, to drive the movement of the lenses in three orthogonal directions with a travel range of 5 mm in each direction. Compared with some other designs that lenses are mounted on the chamber and driven by hand, several advantages are expected: (1) the thermal radiation to the tunneling junction is reduced due to the fact that the lenses stay at $LN_2$ temperature; (2) fine and precise tuning of lens position by piezo actuators improves accuracy of light alignment; (3) hands-free light alignment reduces disturbance in the data acquisition process. Besides, two pairs of windows are used to cover the optical openings on both LHe and $LN_2$ shields to further minimize the thermal radiation to the tunneling junction. In our application, the windows inside the vacuum and the viewports mounted on the SPM chamber along the optical path are all made of fused silica, chosen for its wide range of transmission wavelengths.

Since the LHe vessel in the cryostat is hung by a long tube, it is necessary to fix the LHe shield to the $LN_2$ shield during sample/tip transfer to the SPM scanner head. We designed a clamping mechanism realized by a clamping screw, a lever, a stainless steel belt, and a moving plate with two wedges affixed to it. Clamping/unclamping is achieved by rotating a clamping screw using the wobble stick. During the clamping process, the movements of the related components are indicated by red arrows in Fig. 3 (c). Another magnetic damper is mounted under the LHe shield to dampen the lateral movements between the two shields. The magnets of this damper are glued to the moving plate. A 3D model of the complete cold room, including the shields, the optical components, the clamping structure and the scanner head, is shown in Fig. 3 (d).

### C. SPM scanner head

The scanner head is the most critical unit of an SPM system, as it largely determines the performance of the entire system. Pan-type design,[30-32] due to its rigidity and reliability, proves to be one of the best designs for SPM scanner heads and is used in many systems.[25, 33-35] Figure 4 shows a 3D model of the scanner head we made based on Pan's design. The scanner head is 31 mm × 31 mm × 79 mm in size. As shown in the exploded view of the 3D model of the scanner head in Fig. 4, it is composed of three main parts: the unibody frame, the scanner with the tip-approach module, and the sample stage. This modular design makes it easy to assemble and maintain. The scanner and tip-approach module has the classic structure of the Pan's design: a customized piezo tube is glued coaxially to a sapphire prism, and three pairs of shear piezo stacks drive the sapphire prism in slip-stick mode. A tip holder glued on top of the piezo tube is specially designed to accommodate both the STM tips and the qPlus force sensors, with a compatible design fitting commercially available qPlus force sensors (CreaTec GmbH), providing additional flexibility for the system.[36] The vertical travel distance of the tip is 8 mm.

Unlike Pan's original design, the whole scanner head body is machined out of one piece of titanium by wire-electrode cutting and electric discharge machining to be a unibody frame. Therefore, improved rigidity is expected for the scanner head. We chose titanium as the body material to minimize the thermal drift of the tip, since the thermal expansion coefficient of titanium is close to that of sapphire. A coarse motion mechanism similar to Pan's design is used in the sample stage module to drive the sample's movement in X and Y directions (5 mm in each direction). The sample stage is mounted on top of the unibody frame. A silicon diode temperature sensor and a resistive heater are imbedded in the sample stage module.

Tip-approach failure sometimes happens due to improper compressive force of the Pan's stepper motor, degrading of shear piezo stacks after long-term use, or decreased sensitivity of piezo pieces at low working temperature, and so on. One easy way to address this issue is to adjust the compressive force. In our design, we use one screw at the front side of the scanner head to adjust the compressive force of the Pan's stepper motor. This makes *in situ* adjustment possible using the wobble-stick under vacuum and low temperature conditions.

## II. PERFORMANCE

To test the lowest working temperature of the scanner head, we covered the optical openings on the cold room with aluminum foils. After cooling down, the temperature of the sample is around 5 K and lasts over 70 h. The frequency spectrum of the background current under 5 K with tip retracted is shown in Fig. 5 (a); the highest peak is below 8 fA/√Hz. The frequency spectrum of the tunneling current with feedback off is shown in Fig. 5 (b), and the peaks are far below 1 pA/√Hz. These current frequency spectra indicate that the vibration isolation of the SPM is sufficient and the electrical grounding is effective. To demonstrate the stability of the SPM, we carried out a drift test at 5 K. The lateral drift of the scanner head is 50 pm/h, which is tested by a continuous scanning of 12 h at a fixed area on a Au(111) sample with $I_t$ = 300 pA, $V_b$ = -1 V. The vertical drift of the scanner head is 35 pm/h, which was tested by continuously recording the tip position for 5 h with feedback on, under tunneling conditions with $I_t$ = 500 pA, $V_b$ = -1 V.

To demonstrate the spatial resolution as well as the imaging quality of our system, we used an electrochemically etched tungsten tip to scan cleaned Au(111) substrate. The herringbone structure and atomic-resolution image of Au(111) were clearly resolved by STM, as shown in Figs. 6 (a) and (b). In addition, STS data were obtained on both the fcc and hcp sites, as shown in Fig 6 (c), consistent with the results reported by W. Chen *et al.*[37] During the acquisition of the *dI/dV* spectra, the tip was positioned at a fixed height with tunneling parameters of $I_t$ = 1 nA, $V_b$ = -500 mV, and the modulation voltage amplitude was 20 mV.

To demonstrate the system's capability of *in situ* material growth and characterization, we deposited tellurium (Te) onto the Au(111) substrate by MBE. After annealing, we found that a periodic hexagonal lattice of Te atoms had formed. The atomic structure is clearly resolved by the STM, as shown in Fig. 6 (d), with each bright spot representing a Te atom. The Te atoms are adsorbed onto the √3 sites of the Au(111) lattice. In each unit cell of this hexagonal structure, one triangular patch of half the Te atoms sits on the fcc sites while the other patch sits on the hcp sites. This same structure has been prepared and reported by other groups recently.[38, 39] The growth of the Te hexagonal structure on Au(111) using MBE, together with the clearly-resolved atomic structures, demonstrates the capabilities of this system in both material growth and characterization.

**Conclusions**

We report making a UHV SPM system with MBE and optical access. The design of the system is presented and discussed. The SPM scanner head has a compact, rigid, and modular structure and can work in either STM mode or AFM mode. Optical access to the SPM is implemented in the design of the scanner head, the cold room, and the SPM chamber. At liquid helium temperature, the herringbone reconstruction feature, atomic resolution images, and $dI/dV$ spectra of the Au(111) surface were obtained. The performance of the system is also demonstrated by the preparation and characterization of a periodic structure of Te on Au(111) surface using MBE and STM. This system not only satisfies the basic requirements for material growth and characterization, but is also highly compatible and stable.

**Author Contributions**

H.-J. Gao supervised the project. Q. Huan designed the whole system. Under the guidance of Q. Huan, Z.-B. W. designed the chambers and the scanner head, and Z.-Y. G. designed the clamping mechanism. Z.-B. W. and Z.-Y. G. worked together to set up the system, completed the system tests, and contributed equally to this work. X.-Y. C., Y.-Q. X., and H. Y. assisted in the set-up and testing of the system. All of the authors discussed the construction of the system and joined the writing of the manuscript.

**Acknowledgements**

This work was supported by the National Key Scientific Instrument and Equipment Development Project of China (No. 2013YQ1203451). The authors thank Dr. Christoph Rüdt from CreaTec Fischer & Co. GmbH for his kind help in designing the tip holder.

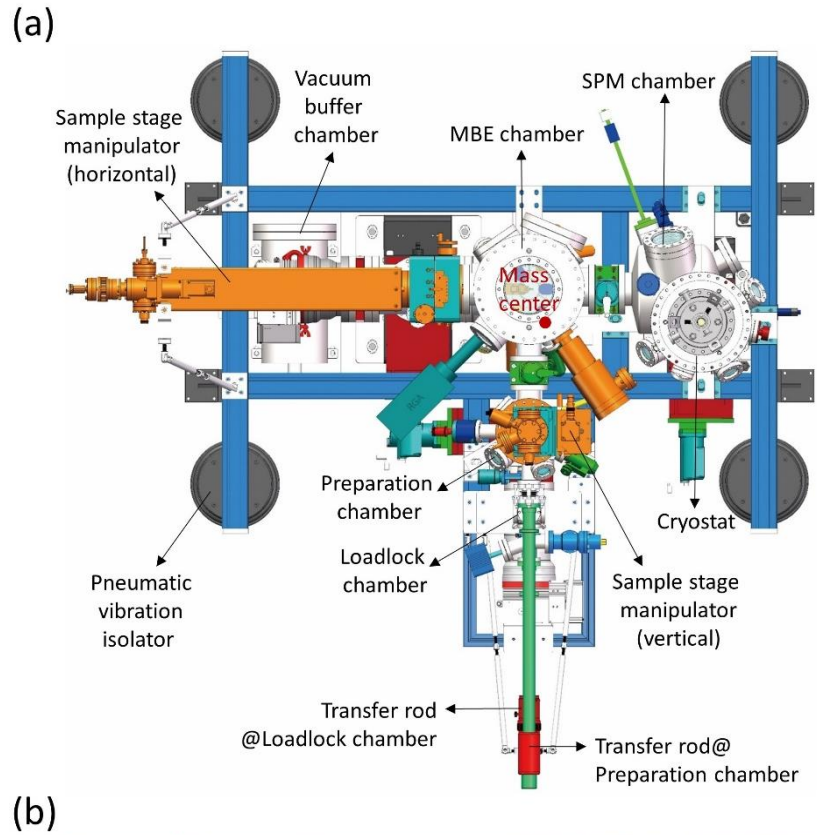

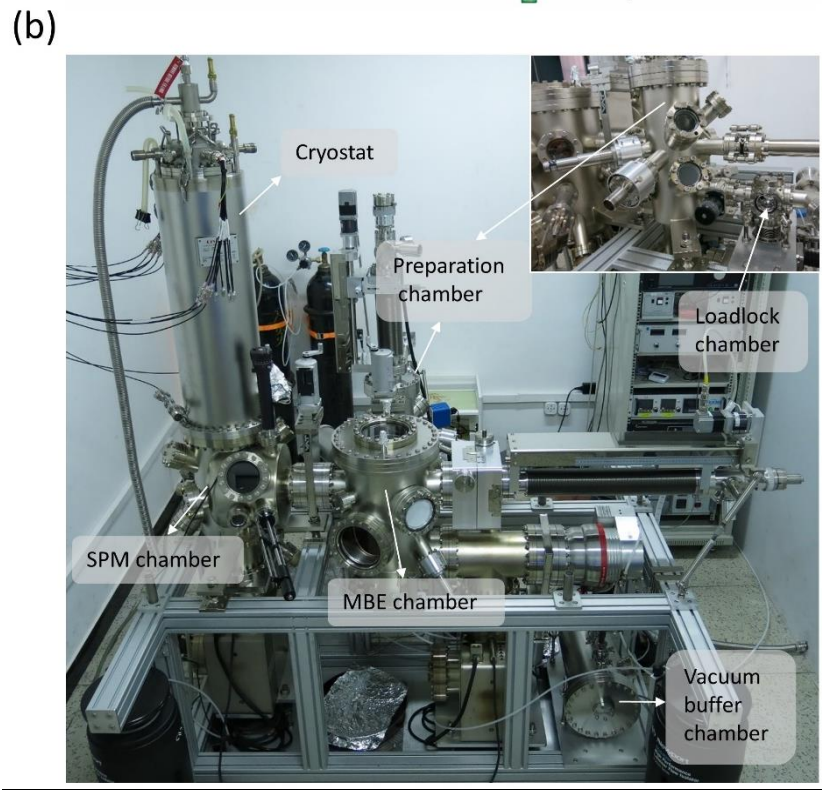

**Fig. 1 3D model and photograph of the system.** (a) Top view of a 3D model of the system. (b) Photograph of the system. The preparation chamber and load-lock chamber are shown in the inset.

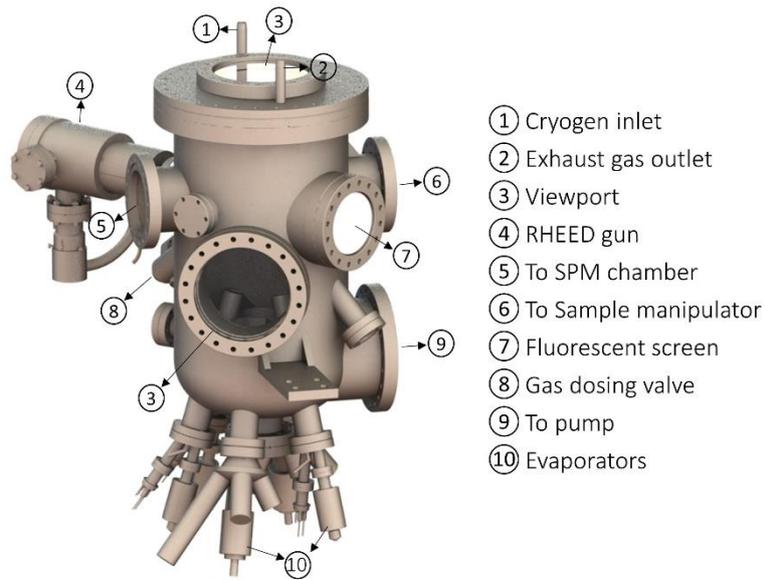

**Fig. 2 3D model of the MBE chamber.** A sample manipulator is installed on the chamber through a DN100 ConFlat flange, as indicated by No. 6 in this model. The MBE chamber connects to the SPM chamber though a port indicated by No. 5 and separated by a gate valve. The preparation chamber connects to a DN75 ConFlat flange at the rear side of this model.

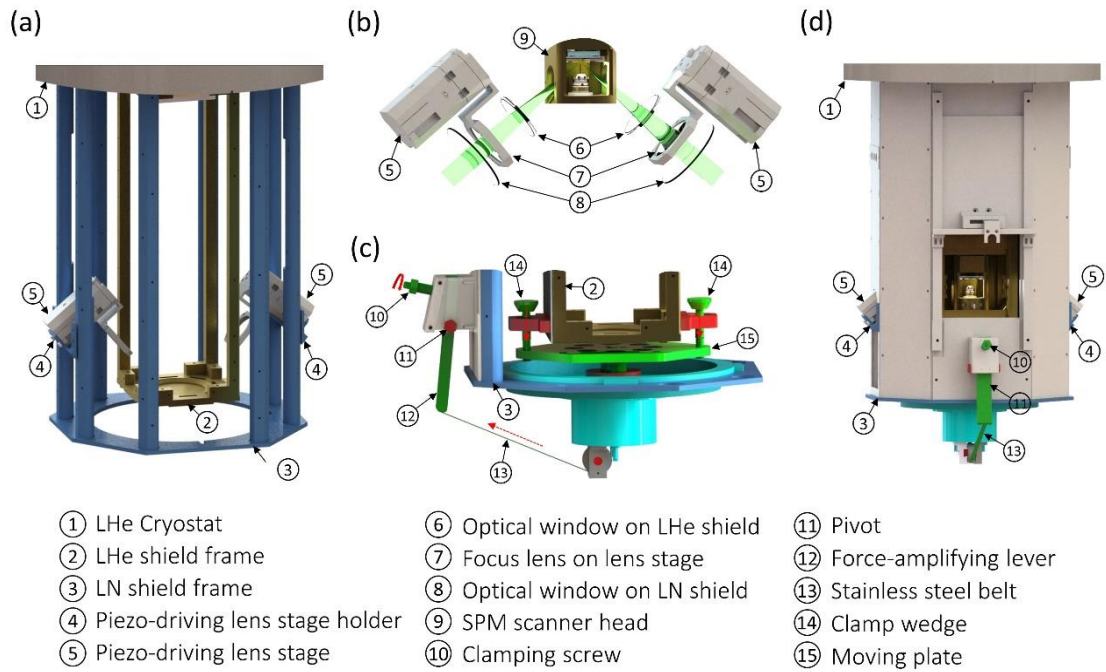

① LHe Cryostat
② LHe shield frame
③ LN shield frame
④ Piezo-driving lens stage holder
⑤ Piezo-driving lens stage
⑥ Optical window on LHe shield
⑦ Focus lens on lens stage
⑧ Optical window on LN shield
⑨ SPM scanner head
⑩ Clamping screw
⑪ Pivot
⑫ Force-amplifying lever
⑬ Stainless steel belt
⑭ Clamp wedge
⑮ Moving plate

**Fig. 3 Structures of the cold room.** (a) The frames of the LHe shield (golden) and the $LN_2$ shield (blue). Two focus lenses are mounted on a pair of piezo-driven lens stages fastened on the $LN_2$ shield. (b) Optical components inside the cold room along the optical path, including two pairs of windows, made of fused silica, mounted on the LHe shield and $LN_2$ shield, and a pair of focus lenses. (c) Clamping structure mounted at the bottom of the $LN_2$ frame to clamp the LHe shield to the $LN_2$ shield. (d) The complete structure of the cold room with the SPM scanner head inside the LHe shield.

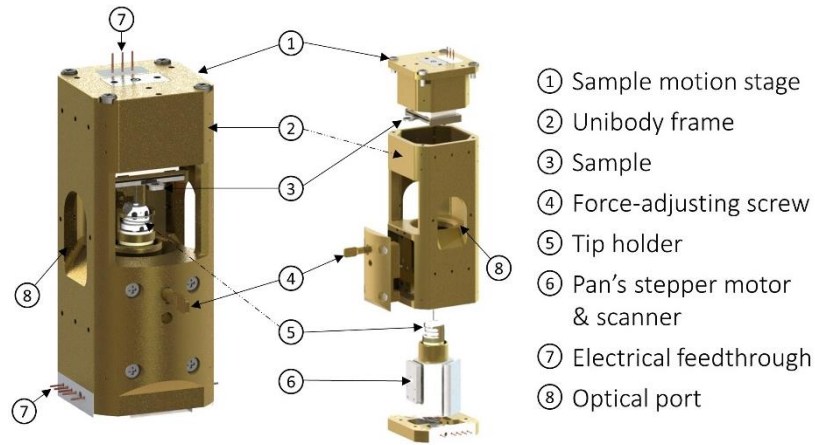

**Fig. 4 3D model and exploded view of the scanner head.** The size of the scanner head is 31 mm × 31 mm × 79 mm. The stepper motor for tip-approach is designed like Pan's, with a working distance of 8 mm. The sample is placed on a sample stage module with a working range of 5 mm × 5 mm. Two optical ports are opened on both sides of the body for optical access to the tunneling junction.

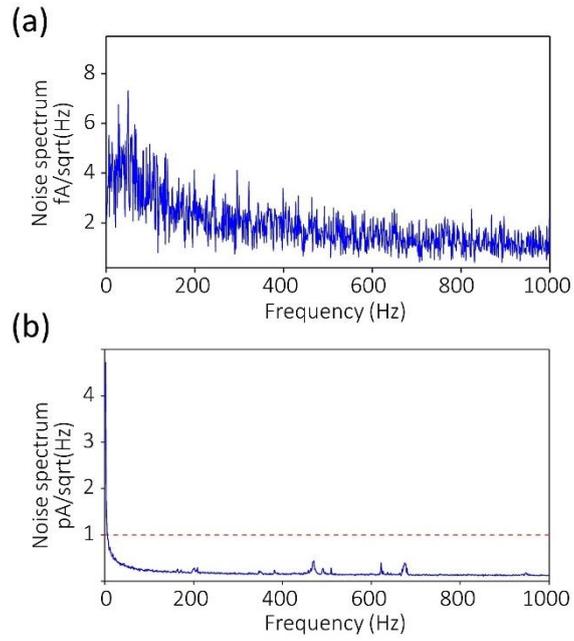

**Fig. 5 Frequency spectra of currents at low temperature.** (a) Frequency spectrum of the background current with tip retracted; (b) Frequency spectrum of the tunneling current with tip approached and feedback turned off. Before the acquisition of this spectrum, the tip is positioned at a fixed height with $I_t$ = -500 pA, $V_b$ = -1 V.

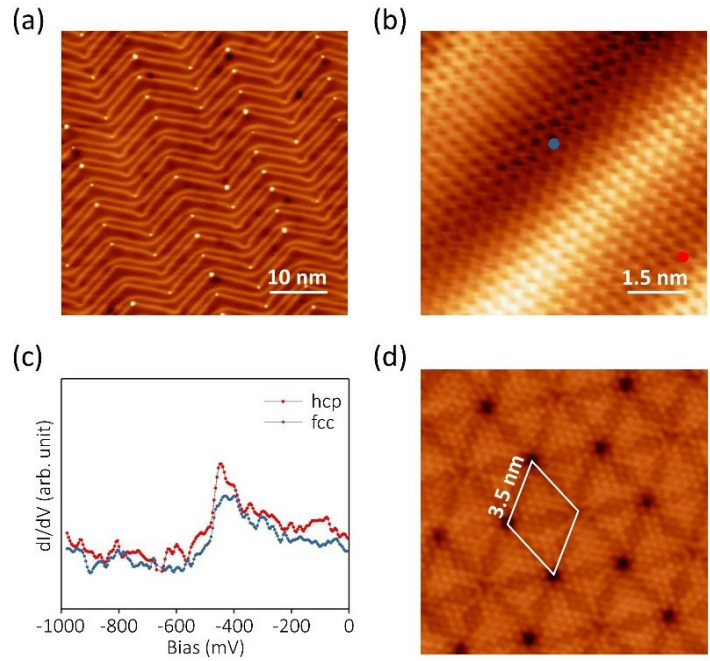

**Fig. 6 STM characterization of Au(111) surface and Te-covered Au(111) surface at 5 K.** (a) Herringbone reconstruction of Au(111) surface, $I_t$ = 0.1 nA, $V_b$ = -1 V. The bright dots at the elbows represent Te atoms. (b) Atomic resolution of the Au(111) surface. $I_t$ = 0.5 nA, $V_b$ = -1 V. (c) The *dI/dV* spectra at the fcc and hcp sites of the Au(111) surface. The hcp site is marked by a red dot and the fcc site is marked by a blue dot in (b). (d) Atomic resolution image of the hexagonal structure of the Te-covered Au(111) surface. $I_t$ = 0.3 nA, $V_b$ = -1 V.